\documentclass{aa}
\usepackage{graphicx}
\usepackage{txfonts}

\begin{document}
\title{
Warm water vapor envelope in the supergiants \mbox{$\alpha$~Ori}\ and \mbox{$\alpha$~Her}\ 
and its effects on the apparent size from the near-infrared 
to the mid-infrared
}

\author{K.~Ohnaka}

\offprints{K.~Ohnaka, \\ \email{kohnaka@mpifr-bonn.mpg.de}}

\institute{
Max-Planck-Institut f\"{u}r Radioastronomie, 
Auf dem H\"{u}gel 69, D-53121 Bonn, Germany
}

\date{Received / Accepted }

\abstract{
We present a possible interpretation for the increase of the 
angular diameter of the supergiants \object{\mbox{$\alpha$~Ori}}\ 
(M1-2 Ia-Ibe) and \object{\mbox{$\alpha$~Her}}\ (M5 Ib-II) 
from the $K$ band to the 11~\mbox{$\mu$m}\ region 
and the high-resolution 11~\mbox{$\mu$m}\ spectra without 
any salient spectral features revealed by 
Weiner et al. (\cite{weiner03a}). 
The angular diameters as well as the high-resolution 
spectra of \mbox{$\alpha$~Ori}\ and \mbox{$\alpha$~Her}\ obtained in the 
11~\mbox{$\mu$m}\ region 
can be reproduced by a warm water vapor envelope, whose presence in 
\mbox{$\alpha$~Ori}\ was revealed by Tsuji (\cite{tsuji00a}) based on 
the reanalysis of the near-infrared data obtained with the Stratoscope II. 
While prominent absorption due to \mbox{H$_2$O}\ can be expected from such a 
dense, warm water vapor envelope, the absorption lines can be filled 
in by emission from the extended part of the envelope.  
This effect leads to a significant weakening of 
the \mbox{H$_2$O}\ lines in the 11~\mbox{$\mu$m}\ region, 
and makes the observed spectra 
appear to be rather featureless and continuum-like.  
However, the emission due to \mbox{H$_2$O}\ lines from the extended 
envelope leads to an increase of the apparent size in this spectral 
region.   The observed angular diameter 
and the high resolution spectra of \mbox{$\alpha$~Ori}\ and 
\mbox{$\alpha$~Her}\ in the 11~\mbox{$\mu$m}\ region can be 
best interpreted by the water vapor envelope extending to 
1.4 -- 1.5~\mbox{$R_{\star}$}, with a temperature of $\sim 2000$~K and 
a column density of \mbox{H$_2$O}\ of the order of 
$10^{20}$~\mbox{cm$^{-2}$}.  
}   

\titlerunning{Warm water vapor envelope in \mbox{$\alpha$~Ori}\ and 
\mbox{$\alpha$~Her}}
\authorrunning{K.~Ohnaka}
\maketitle

\section{Introduction}
\label{sect_intro}

Recent mid-infrared interferometric observations have revealed that the 
diameters of late-type stars such as M supergiants and Mira variables 
increase from the $K$ band to the $N$ band.  
Weiner et al. (\cite{weiner00}, \cite{weiner03a}, hereafter W00 and 
WHT03a, respectively) observed the supergiants \mbox{$\alpha$~Ori}\ 
(M1-2 Ia-Ibe) and \mbox{$\alpha$~Her}\ (M5 Ib-II) 
at 11~\mbox{$\mu$m}, using the Infrared Spatial Interferometer (ISI) 
with a narrow spectral bandwidth of 0.17~\mbox{cm$^{-1}$}.  They found that 
the uniform disk (UD) diameters of these stars are $\sim$ 30\% 
larger than those measured in the $K$ band, which at least for these stars 
are considered to approximately represent the continuum radii, 
as compared to those measured in the optical where scattering due to 
dust particles is more pronounced (see, e.g., discussion in 
Dyck et al. \cite{dyck92}).  W00 and WHT03a show that 
the angular diameter of \mbox{$\alpha$~Ori}\ at 11~\mbox{$\mu$m}\ 
ranges from 53 to 56~mas (43 -- 44~mas in the $K$ band derived by 
Dyck et al. \cite{dyck92} and Perrin et al. \cite{perrin04}), 
while that of \mbox{$\alpha$~Her}\ at 11~\mbox{$\mu$m}\ is 39~mas 
(31 -- 32~mas in the $K$ band derived by Benson et al. \cite{benson91} and 
Perrin et al. \cite{perrin04}).  

Although dust emission is detected for \mbox{$\alpha$~Ori}, 
the extended dust shell is unlikely to be responsible for this increase 
of the diameter for the following reason. 
The visibility expected for a system consisting of a stellar disk 
and a very extended dust shell is characterized by a steep 
drop at low spatial frequencies and a gradual decrease at higher 
spatial frequencies.  The amount of the steep drop at low spatial 
frequencies corresponds to the flux contribution of the extended 
dust shell in the field of view, while the stellar disk affects 
the visibility shape at higher spatial frequencies.  
The inner radius of the dust shell around M supergiants such as 
\mbox{$\alpha$~Ori}\ is derived to be $\sim 50$~\mbox{$R_{\star}$}\ 
by Danchi et al. (\cite{danchi94}).  
Such a large dust shell is completely resolved with the baselines 
used by W00 and WHT03a (20 -- 56~m), that is, 
the dust shell does not affect the shape of the observed visibility 
function at these baselines.  
The presence of the 
dust shell lowers the total visibility at these baselines 
by an amount equal to the fraction of flux coming from the dust shell 
in the field of view.  
W00 and WHT03a take this effect into account in 
deriving the uniform disk diameters, and 
show that the observed visibilities can be well fitted by a uniform 
disk, but with diameters clearly larger than those measured in the 
$K$ band.  
For \mbox{$\alpha$~Her}, Danchi et al. (\cite{danchi94}) detected a 
circumstellar dust shell, but more recent spectrophotometric observations by 
Monnier et al. (\cite{monnier98}) show that \mbox{$\alpha$~Her}\ exhibits very 
little dust emission.  WHT03a suggest that the dust shell around 
\mbox{$\alpha$~Her}\ might have evolved since the observations by 
Danchi et al. (\cite{danchi94}).  Therefore, the increase of 
the angular diameter observed in \mbox{$\alpha$~Ori}\ and 
\mbox{$\alpha$~Her}\ cannot simply be 
attributed to the dust shell, while it cannot be completely ruled
out that a possible presence of small dust clumps close to the star 
affects the observed visibility, as Bester et al. (\cite{bester96}) 
suggest based on their ISI observations of \mbox{$\alpha$~Ori}.

WHT03a also present high-resolution spectra of \mbox{$\alpha$~Ori}\ and 
\mbox{$\alpha$~Her}\ 
in the same wavelength range as they selected for the ISI observations. 
These high-resolution spectra were obtained with 
the TEXES instrument mounted on the Infrared Telescope Facility 
with a spectral resolution of $\sim 10^5$ (Lacy et al. \cite{lacy02}).  
A glance of these TEXES high-resolution spectra (Fig.~2 of WHT03a) 
reveals that neither \mbox{$\alpha$~Ori}\ nor \mbox{$\alpha$~Her}\  
shows any significant spectral features in the bandpasses 
used for the ISI observations.  WHT03a conclude 
that the interferometric observations with ISI were 
carried out in the bandpasses which exhibit continuous 
spectra free from any molecular and/or atomic features.  
That is, the diameters measured in the 11~\mbox{$\mu$m}\ region which 
appears to be the continuum are significantly larger than the 
$K$ band diameters, which are also considered to be rather close 
to the diameter measured in the continuum.  Therefore, 
the interferometric and spectroscopic observations by WHT03a 
pose a serious problem in interpreting the increase of the angular diameter 
from the near-infrared to the mid-infrared.  
As possible explanations for these apparently inconsistent results, 
WHT03a suggest that the density stratification may be very 
nonhydrostatic near the photosphere and/or that the angular diameters 
measured in the near-infrared may be affected by a presence of hot
spots, which may make the apparent size of a star smaller than it is. 

Tsuji (\cite{tsuji00a}, hereafter T00a) 
discovered a warm water vapor layer around two supergiants 
\mbox{$\alpha$~Ori}\ and $\mu$~Cep based on the reanalysis of the 
near-infrared spectra obtained with the Stratoscope II.  
The presence of water 
vapor in \mbox{$\alpha$~Ori}\ was further confirmed by the detection of weak 
\mbox{H$_2$O}\ absorption in the 6~\mbox{$\mu$m}\ spectrum obtained with the 
Infrared Space Observatory (ISO) (Tsuji \cite{tsuji00b}). 
T00a also studied the pure-rotation 
lines of \mbox{H$_2$O}\ identified in the high-resolution spectrum of 
\mbox{$\alpha$~Ori}\ at 12 \mbox{$\mu$m}\ obtained by 
Jennings \& Sada (\cite{jennings98}).  
Although the observed equivalent widths of the water lines at 
12~\mbox{$\mu$m}\ are larger than those predicted from the photospheric 
model of T00a by a factor of 2 -- 3, 
the 12 \mbox{$\mu$m}\ water lines are expected to be even stronger than 
observed, if the column density of water molecules derived from 
the near-infrared spectrum is correct.  T00a and Tsuji (\cite{tsuji00b}) 
point out the possibility that the absorption may be filled in by 
emission from the extended part of the water vapor envelope.  

If the warm water vapor envelope is present and the emission from it 
can make the 12 \mbox{$\mu$m}\ \mbox{H$_2$O}\ lines appear to be weaker 
by the filling-in 
effect, it may also be the case for the 11~\mbox{$\mu$m}\ lines presented in 
WHT03a.  That is, the absorption is filled in by the emission resulting 
from the geometrical extension of the warm molecular envelope, 
making the spectra almost featureless.  While such featureless 
mid-infrared spectra may not show any hint of the presence of the warm 
molecular envelope 
and manifest themselves as the continuum, the extended, warm molecular 
envelope can cause the apparent diameter in the mid-infrared to be larger 
than the photospheric diameter.  
Therefore, the warm water vapor envelope 
may explain the increase of angular diameters from the $K$ band to the 
$N$ band, simultaneously with the spectra observed in the 
11~\mbox{$\mu$m}\ region. 

In the present paper, we examine this possibility for 
\mbox{$\alpha$~Ori}\ and \mbox{$\alpha$~Her}.  
We compare angular diameters as well as spectra predicted from 
a simple model for the warm water vapor envelope 
with several observational data: 11~\mbox{$\mu$m}\ and 
12~\mbox{$\mu$m}\ high-resolution 
spectra, the ISO spectrum in the 6~\mbox{$\mu$m}\ region, and 
the angular diameters measured at 11~\mbox{$\mu$m}.

\section{Modeling of the warm molecular envelope}

\begin{figure}
\begin{center}
\resizebox{7.5cm}{!}{\rotatebox{0}{\includegraphics{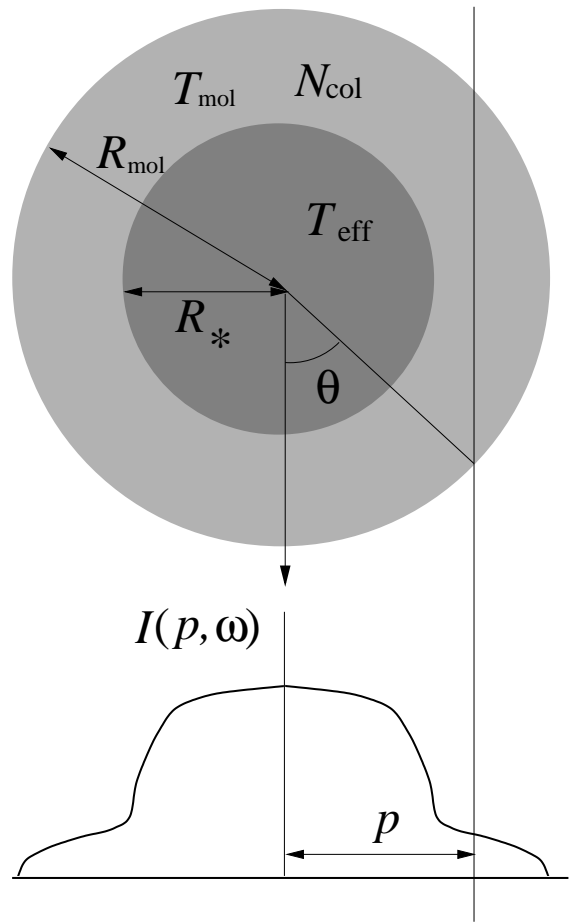}}}
\end{center}
\caption{The model used in the present work.   The photosphere is 
approximated with a blackbody of \mbox{$T_{\rm eff}$}, with a radius 
\mbox{$R_{\star}$}.  
The column density in the radial direction (\mbox{$N_{\rm col}$}) and 
the temperature of 
\mbox{H$_2$O}\ gas (\mbox{$T_{\rm mol}$}) as well as the radius of the 
water vapor envelope (\mbox{$R_{\rm mol}$}) are the input parameters.  
}
\label{model}
\end{figure}

Our model consists of the central star, which is represented by 
a blackbody of effective temperature \mbox{$T_{\rm eff}$}, and a warm 
molecular 
envelope extending to \mbox{$R_{\rm mol}$}, as depicted in Fig.~\ref{model}. 
As the figure shows, 
the inner radius of the molecular envelope is set to be equal to 
the radius of the star \mbox{$R_{\star}$}. 
We adopt an effective temperature of 3600~K for \mbox{$\alpha$~Ori}\ (Tsuji et
al. \cite{tsuji94}) and 3200~K for \mbox{$\alpha$~Her}\ 
(Tsuji \cite{tsuji81}). 
The input parameters are the temperature of water vapor gas 
($T_{\rm mol}$), its column density in the radial direction 
(\mbox{$N_{\rm col}$}), 
and the geometrical extension of the envelope 
($R_{\rm mol}$ in units of $R_{\star}$).  The temperature and the 
density of the water vapor are assumed to be constant in the 
envelope.  

We first calculate the line opacity due to \mbox{H$_2$O}\ 
using the HITEMP line list (Rothman \cite{rothman97}), with a Gaussian 
profile assumed.  We adopt a velocity of 5~\mbox{km s$^{-1}$}\ for the
sum of the 
thermal velocity and the microturbulent velocity.  
The energy level populations of \mbox{H$_2$O}\ are calculated in local 
thermodynamical equilibrium (LTE).  The validity of LTE can be 
examined, using order-of-magnitude estimates of collisional and 
radiative de-excitation rates, as adopted by Ryde et al. (\cite{ryde02}). 
The collisional de-excitation rate $C_{\rm ul}$ is given by 
$C_{\rm ul} \sim N_{\rm H} \sigma_{\rm ul} v_{\rm rel}$, 
where $N_{\rm H}$ is the density of H atoms, which 
are assumed to be the primary collision partner, $\sigma_{\rm ul}$ 
is the cross section, which we approximate with the geometrical 
cross section, and $v_{\rm rel}$ is the 
relative velocity between the H atoms and \mbox{H$_2$O}\ molecules.  
As we will show below, the column densities of \mbox{H$_2$O}\ of the warm 
water vapor envelope in \mbox{$\alpha$~Ori}\ and \mbox{$\alpha$~Her}\
are derived to be 
2 -- $7 \times 10^{20}$~\mbox{cm$^{-2}$}.  The radius of the water vapor 
envelope is derived to be 1.4 -- 1.5~\mbox{$R_{\star}$}, which is translated 
into $\sim 2 \times 10^{13}$~cm with a stellar radius of 
650~\mbox{$R_{\sun}$}\ 
assumed.  The number density of \mbox{H$_2$O}\ is then estimated to be 
1.0 -- $4 \times 10^{7}$~\mbox{cm$^{-3}$}.  The ratio of the number 
density of H atoms to that of \mbox{H$_2$O}\ molecules expected 
in chemical equilibrium is approximately $10^{4}$ -- $10^{5}$ 
for the relevant temperatures and densities.  Therefore, the number 
density of H atoms is estimated to be 
$1 \times 10^{11}$ -- $4 \times 10^{12}$~\mbox{cm$^{-3}$}.  
With a geometrical 
cross section $\sigma_{\rm ul}$ of $10^{-15}$~cm$^2$ and a relative 
velocity $v_{\rm rel}$ of 5~\mbox{km s$^{-1}$}\ assumed, these number 
densities of H atoms lead to collisional de-excitation rates of 
$\sim 50$ -- 2000~s$^{-1}$.  
On the other hand, the rate of spontaneous emission can be 
estimated from the Einstein coefficients $A_{\rm ul}$.  
For the \mbox{H$_2$O}\ molecule, the ranges of $A_{\rm ul}$ are approximately 
$\sim 10^{-4}$ -- $3 \times 10^{2}$~s$^{-1}$, 
$\sim 10^{-5}$ -- $3 \times 10^{2}$~s$^{-1}$, and 
$\sim 10$~s$^{-1}$ for the 11 -- 12~\mbox{$\mu$m}\ region, 
$K$ band, and \mbox{$L^{\prime}$}\ band, respectively.  Therefore, 
for the wavelength regions that we will discuss below, 
the assumption of LTE is valid for weak and moderately strong
\mbox{H$_2$O}\ lines, 
while non-LTE effects may not be negligible for strong lines.  However, 
a quantitative assessment of non-LTE effects is beyond the scope 
of the present paper, and we assume LTE for the \mbox{H$_2$O}\ lines 
considered here. 

Once the line 
opacity is calculated, the intensity distribution at the wavenumber 
$\omega$ can be calculated as 
\begin{equation}
 I(p, \omega) = B_{\omega}(\mbox{$T_{\rm mol}$})(1 - e^{-\tau_{\omega}}) + 
 B_{\omega}(\mbox{$T_{\rm eff}$}) e^{-\tau_{\omega}} \, 
{\rm circ}(p/\mbox{$R_{\star}$}),
\end{equation}
where $p$ is the impact parameter, 
$\tau_{\omega}$ is the optical depth along the ray at $p$, 
and ${\rm circ}(p/\mbox{$R_{\star}$})$ takes a value of 1 for 
$p < \mbox{$R_{\star}$}$ and 0 elsewhere (see Fig.~\ref{model}). 
In order to obtain the spectrum, the emergent flux at each 
wavenumber, $F_{\omega}$, is calculated by 
\begin{equation}
 F_{\omega} = 2 \pi \int_{0}^{1} I(p, \omega) \, \mu \, d\mu,
\label{fluxInteg}
\end{equation}
where $\mu$ is defined as 
$\mu = \cos \theta = \sqrt{1 - (p/\mbox{$R_{\rm mol}$})^2}$ 
with $\theta$ defined in Fig.~\ref{model}.  
With an appropriate wavenumber interval, we calculate an 
intensity profile and a flux at each wavenumber.  The corresponding 
(monochromatic) visibility is calculated from this intensity profile 
using the Hankel transform (two-dimensional Fourier transform for 
axisymmetric objects).

\section{Comparison with the observed spectra and angular diameters:
\mbox{\boldmath $\alpha$}~Ori}

Using the above model for the warm \mbox{H$_2$O}\ envelope, 
we compare the synthetic spectra and the visibilities with 
the observed data, 
which include the 11~\mbox{$\mu$m}\ spectra presented in WHT03a, 
the 12~\mbox{$\mu$m}\ spectra obtained by 
Jennings \& Sada (\cite{jennings98}),  
the 6~\mbox{$\mu$m}\ spectrum in Tsuji (\cite{tsuji00b}), and the 
interferometric observations at 11~\mbox{$\mu$m}\ obtained by W00 and WHT03a.  
We change the input parameters (\mbox{$T_{\rm mol}$}, 
\mbox{$R_{\rm mol}$}, \mbox{$N_{\rm col}$}), 
and search for the combination which can simultaneously reproduce 
these observational results.  
The range of the input parameters used in the calculations is 
as follows: \mbox{$T_{\rm mol}$}\ (K) = 2150, 2100, 2050, 2000, 1950,
1900, 1800, 1700, 1600, 1400, and 1200, 
\mbox{$N_{\rm col}$}\ (\mbox{cm$^{-2}$}) = $1 \times 10^{19}$, 
$1 \times 10^{20}$, $2 \times 10^{20}$, $5 \times 10^{20}$, 
$1 \times 10^{21}$, and $1 \times 10^{22}$, 
\mbox{$R_{\rm mol}$}\ (\mbox{$R_{\star}$}) = 1.3, 1.4, 1.45, 1.5, 1.55, 
1.6, and 1.7.  It has turned out 
that the aforementioned observational data for \mbox{$\alpha$~Ori}\ can 
be best reproduced with \mbox{$T_{\rm mol}$}\ = 2050~K, 
\mbox{$R_{\rm mol}$}\ = 1.45~\mbox{$R_{\star}$}, 
$N_{\rm col} = 2 \times 10^{20}$~\mbox{cm$^{-2}$}.  
The uncertainties of these values are estimated by changing 
the parameters around the best-fit parameter set by small amounts. 
We estimate that the uncertainties of 
the gas temperature, the radius, and the column density of \mbox{H$_2$O}\ 
are $\pm 100$~K, $\pm 0.1$~\mbox{$R_{\star}$}, and a factor of $\sim 2$, 
respectively.  The gas temperature and the column density of 
\mbox{H$_2$O}\ molecules derived in the present work are in rough agreement 
with those derived by T00a.  
In the following subsections, we will discuss 
the comparison for each observational data set.

\subsection{11~\mbox{$\mu$m}\ spectra}
\label{sect_alfOri11mu}

\begin{figure}
\begin{center}
\resizebox{8.5cm}{!}{\rotatebox{-90}{\includegraphics{0668fig2.ps}}}
\end{center}
\caption{Spectra in the 11~\mbox{$\mu$m}\ region. 
The dots represent the spectra of \mbox{$\alpha$~Ori}\ presented in 
WHT03a, while 
the solid lines represent the calculated spectra from the best-fit 
model with \mbox{$T_{\rm mol}$}\ = 2050~K, 
\mbox{$R_{\rm mol}$}\ = 1.45~\mbox{$R_{\star}$}, 
and \mbox{$N_{\rm col}$}\  = $2 \times 10^{20}$~\mbox{cm$^{-2}$}. 
The synthetic spectra are convolved with a Gaussian with a FWHM of 
0.037~\mbox{cm$^{-1}$}\ to account for the effects of the instrument as well 
as of the macroturbulent velocity, and redshifted by 39~\mbox{km s$^{-1}$}\ to 
match the observations.  The positions of the \mbox{H$_2$O}\ lines whose 
intensity at 2000~K is stronger than 
$3 \times 10^{-23}$~cm molecule$^{-1}$ are marked with ticks. 
These line positions are taken from the HITEMP database, and 
are also redshifted by 39~\mbox{km s$^{-1}$}\ with respect to the rest 
wavenumber. 
}
\label{alfOriSp11mu}
\end{figure}

\begin{figure}
\begin{center}
\resizebox{8.5cm}{!}{\rotatebox{-90}{\includegraphics{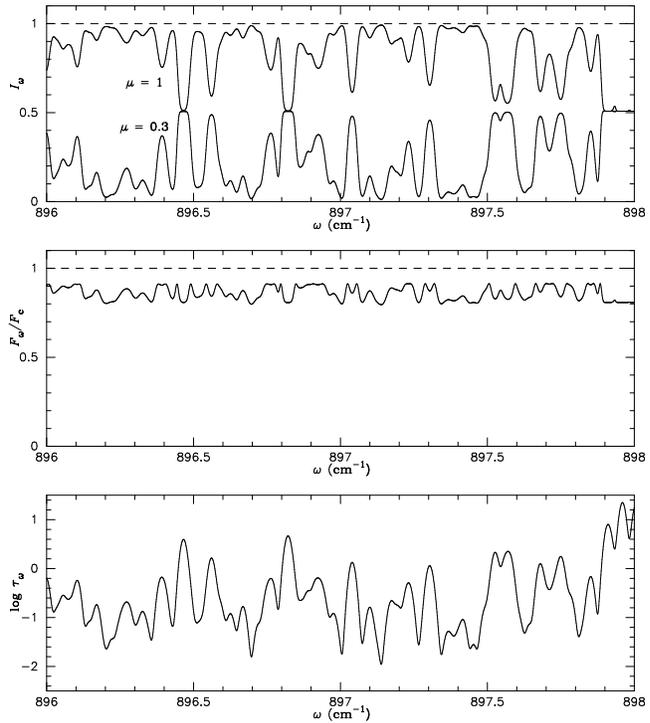}}}
\end{center}
\caption{{\bf Top panel}: Spectra predicted from the best-fit 
model for \mbox{$\alpha$~Ori}\ at the stellar disk center 
($\mu = 1$, upper spectrum) and near the limb of the warm \mbox{H$_2$O}\ 
envelope ($\mu = 0.3$, lower spectrum) in the region around 
11.1494~\mbox{$\mu$m}.  The continuum 
level is shown with the dashed line.  
The flux contribution of dust emission is not included in the spectra shown.  
{\bf Middle panel}: The resulting spectrum (emergent flux) expected 
for the whole object (stellar disk + warm water vapor envelope), 
calculated from the intensity spectra such as shown in the upper panel, 
using Eq. (\ref{fluxInteg}). 
{\bf Bottom panel}: Optical depth 
in the radial direction due to \mbox{H$_2$O}\ lines in the same spectral 
region.  
The spectra and the flux shown in the top and middle panels 
as well as the optical depth shown in the bottom panel are 
redshifted by 39~\mbox{km s$^{-1}$}\ with respect to the rest wavenumber to 
match the observed spectra shown in Fig.~\ref{alfOriSp11mu}, 
but {\em not} convolved with the Gaussian representing the 
instrumental effect and the macroturbulent velocity. 
}
\label{alfOriIntensSp11mu}
\end{figure}

We calculate synthetic spectra in the 11~\mbox{$\mu$m}\ region 
with a wavenumber interval of 
0.001~\mbox{cm$^{-1}$}, and then convolve with a Gaussian profile which 
represents the effects of the instrument and 
the macroturbulent velocity.  
The spectral resolution of the TEXES instrument is $10^5$, 
which translates into an instrumental broadening of 3~\mbox{km s$^{-1}$}.   
Jennings \& Sada (\cite{jennings98}) derived a macroturbulent 
velocity of 12~\mbox{km s$^{-1}$}\ for \mbox{$\alpha$~Ori}.  
Thus, the synthetic spectra are convolved with a Gaussian with a FWHM 
of $\sqrt{3^2 + 12^2} = 12.4$~\mbox{km s$^{-1}$}, 
which corresponds to 0.037~\mbox{cm$^{-1}$}\ at 897~\mbox{cm$^{-1}$}. 
In order to take the continuous dust emission from the circumstellar 
dust shell into account, the convolved and normalized spectrum 
$F_{\omega}$ is diluted as follows:
\begin{equation}
 F_{\omega}^{\rm diluted} = (1 - f_{\rm dust})F_{\omega} + f_{\rm dust} \, ,
\end{equation}
where $F_{\omega}^{\rm diluted}$ is the final spectrum, and $f_{\rm dust}$ 
is the fraction of the flux contribution of the circumstellar dust 
shell.   WHT03a derived $f_{\rm dust} \sim 0.44$ for 
\mbox{$\alpha$~Ori}\ at 11~\mbox{$\mu$m}, and we adopt this value.  

Figure~\ref{alfOriSp11mu} shows a comparison between the synthetic 
11~\mbox{$\mu$m}\ spectra from the best-fit model for 
\mbox{$\alpha$~Ori}\ and the observed 
spectra presented in WHT03a. 
The dots in the figure represent the observed data, which were read 
off Fig.~2 in WHT03a, 
while the solid lines represent the synthetic spectra, 
which are redshifted by 39~\mbox{km s$^{-1}$}\ to match the observations 
(see WHT03a).  
Since our model is too simple to deal with the complicated physical 
processes in the outer atmosphere, we cannot expect very good
quantitative agreement for individual spectral features.  
However, Fig.~\ref{alfOriSp11mu} illustrates that the model calculation 
can reproduce the absence of conspicuous spectral features 
as well as the depths of the small features.  
Although a number of pure-rotation lines of \mbox{H$_2$O}\ are present 
as shown with the ticks in the figure, they appear neither as absorption nor 
as emission.  The reason for the absence of salient spectral features is 
that the absorption lines 
are filled in by the emission from the outer part of the envelope.  
In order to illustrate this effect, we show synthetic spectra 
at the disk center ($\mu = 1$) and near the limb 
of the warm \mbox{H$_2$O}\ envelope ($\mu = 0.3$) 
in Fig.~\ref{alfOriIntensSp11mu} (top panel), 
together with the resulting spectrum (emergent flux) 
expected for the whole system (middle panel), and 
the optical depth due to \mbox{H$_2$O}\ in the radial 
direction (bottom panel).  
As the bottom panel shows, the optical depth due to \mbox{H$_2$O}\ ranges 
mostly from $\sim 0.1$ to $\sim 1$ with several peaks reaching 
$\sim 10$, which means that the warm water vapor envelope is not 
optically thick.  
In such a case, the observer see the 
water vapor envelope with the hotter star in the background 
in a line of sight with $p < \mbox{$R_{\star}$}$, which 
leads to appearance of absorption lines.  
The spectrum at the disk center shown in the top panel illustrates 
this case, where absorption features are observed at the positions 
of the \mbox{H$_2$O}\ lines. 
On the other hand, in a line of sight with 
$\mbox{$R_{\star}$} \le p \le \mbox{$R_{\rm mol}$}$, 
emission lines are observed, as the spectrum near the limb ($\mu = 0.3$) 
illustrates.  
Since the spectrum (i.e. emergent flux) observed for the object 
(central star + water vapor envelope) is obtained by integrating 
the intensity over the area that the object projects onto the 
plane of the sky, the absorption expected from the region with 
$p < \mbox{$R_{\star}$}$ and emission expected from the outer region with 
$\mbox{$R_{\star}$} \le p \le \mbox{$R_{\rm mol}$}$ almost cancel out 
in the resulting spectrum, 
leading to the nearly featureless spectra as shown in the middle panel.

\subsection{12~\mbox{$\mu$m}\ spectra}

\begin{figure}
\begin{center}
\resizebox{8.5cm}{!}{\rotatebox{-90}{\includegraphics{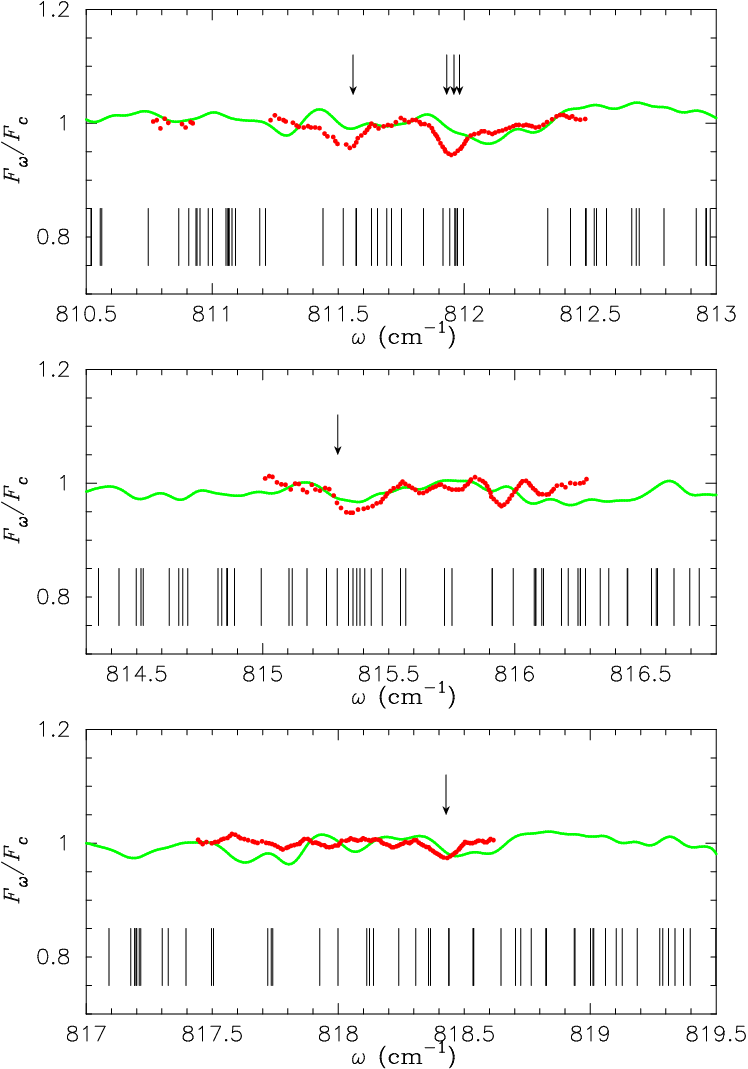}}}
\end{center}
\caption{Spectra in the 12~\mbox{$\mu$m}\ region. 
The dots represent the spectra of \mbox{$\alpha$~Ori}\ observed by 
Jennings \& Sada (\cite{jennings98}), while 
the solid lines represent the calculated spectra of the best-fit 
model.  
The synthetic spectra are 
convolved with a Gaussian with a FWHM of 0.1~\mbox{cm$^{-1}$}\ to match the 
resolution of the spectrometer used by 
Jennings \& Sada (\cite{jennings98}).  
The ticks represent the positions of \mbox{H$_2$O}\ lines whose intensity 
at 2000~K is stronger than $3 \times 10^{-23}$~cm molecule$^{-1}$. 
The observed and predicted spectra as well as the \mbox{H$_2$O}\ 
line positions 
are plotted in the laboratory frame.   The \mbox{H$_2$O}\ lines identified 
by Jennings \& Sada (\cite{jennings98}) are shown with the arrows. 
}
\label{alfOriSp12mu}
\end{figure}

Figure~\ref{alfOriSp12mu} shows the synthetic and observed 
spectra in the 12~\mbox{$\mu$m}\ region.  
The observed spectra, which 
were read off Fig.~1 in Jennings \& Sada (\cite{jennings98}), are 
represented with the dots.  
The pure-rotation lines of \mbox{H$_2$O}\ identified by 
Jennings \& Sada (\cite{jennings98}) are marked with the 
arrows in the figure.  
The synthetic spectra, which are represented with the solid lines, were 
calculated with a wavenumber interval of 0.001~\mbox{cm$^{-1}$}, and 
convolved with a Gaussian with a FWHM of 0.1~\mbox{cm$^{-1}$}, 
which corresponds 
to the resolution of the spectrometer used by 
Jennings \& Sada (\cite{jennings98}).  
As in the case of the 11~\mbox{$\mu$m}\ spectra discussed above, it is 
necessary to include the dust emission from the circumstellar 
dust shell in the synthetic spectra.  
The flux contribution of the dust shell is estimated 
to be 35\% at 12~\mbox{$\mu$m}\ by Jennings \& Sada (\cite{jennings98}).  
Therefore, the calculated spectra are diluted with $f_{\rm dust} = 0.35$.  
Figure~\ref{alfOriSp12mu} shows that the synthetic spectra 
can reproduce the absence of strong spectral features, 
and the observed depths of the weak features are also 
reproduced to some extent.  
The filling-in effect due to the outer part 
of the warm \mbox{H$_2$O}\ envelope weakens the absorption lines 
significantly, as in the case of the 11~\mbox{$\mu$m}\ spectra.  
The absorption features observed at 
815.4~\mbox{cm$^{-1}$}\ and 815.95~\mbox{cm$^{-1}$}\ are affected 
by the blend of 
OH lines, which are not included in the present calculation of the 
synthetic spectra.  It should also be noted that T00a  
shows that the 12~\mbox{$\mu$m}\ \mbox{H$_2$O}\ lines may partially 
originate in 
the photosphere.   If the detailed photospheric structure were 
incorporated in our model, the \mbox{H$_2$O}\ absorption lines would 
be stronger 
than shown in Fig.~\ref{alfOriSp12mu}, which might improve the agreement 
with the observed spectra.

\subsection{6~\mbox{$\mu$m}\ ISO spectrum}

\begin{figure}
\begin{center}
\resizebox{8.5cm}{!}{\rotatebox{-90}{\includegraphics{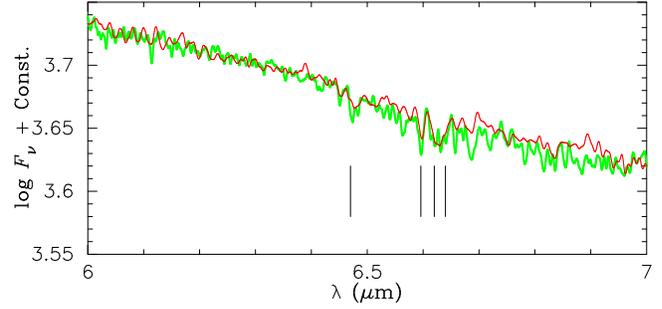}}}
\end{center}
\caption{Spectra in the 6~\mbox{$\mu$m}\ region.  The thin solid line 
represents the 
ISO SWS spectrum of \mbox{$\alpha$~Ori}, while the thick solid line 
represents the 
synthetic spectrum predicted from the best-fit model.  The synthetic 
spectrum is convolved with a Gaussian with a FWHM of 1.0~\mbox{cm$^{-1}$}.  
The ticks mark the \mbox{H$_2$O}\ absorption features identified by 
Tsuji (\cite{tsuji00b}).  
}
\label{alfOriSp6mu}
\end{figure}

Figure~\ref{alfOriSp6mu} shows a comparison between synthetic spectra 
and the spectrum obtained with the ISO Short Wavelength Spectrometer (SWS) 
in the 6 -- 7~\mbox{$\mu$m}\ region where weak absorption due to 
the \mbox{H$_2$O}\ $\nu_2$ fundamental bands was identified by 
Tsuji (\cite{tsuji00b}). 
The thin solid line represents the spectrum of 
\mbox{$\alpha$~Ori}\ (observed on 
1997 October 8 UT) retrieved from the ISO data archive, while 
the synthetic spectrum from the best-fit model, which is convolved with 
a Gaussian with a FWHM of 1.0~\mbox{cm$^{-1}$}\ to match the resolution 
of the ISO SWS spectrum ($R \simeq 1600$), 
is shown with the thick solid line.  
No dilution due to the circumstellar dust emission is assumed in 
this wavelength region.  
Figure~\ref{alfOriSp6mu} demonstrates that the model can fairly reproduce 
the \mbox{H$_2$O}\ absorption features observed with ISO SWS, which are marked 
with the ticks in the figure.  
Therefore, we conclude that our warm \mbox{H$_2$O}\ envelope model is also 
consistent with the 6~\mbox{$\mu$m}\ \mbox{H$_2$O}\ spectrum of 
\mbox{$\alpha$~Ori}\ observed with ISO SWS.

\subsection{Angular diameter in the 11~\mbox{$\mu$m}\ region}

\begin{figure}
\begin{center}
\resizebox{8.2cm}{!}{\rotatebox{-90}{\includegraphics{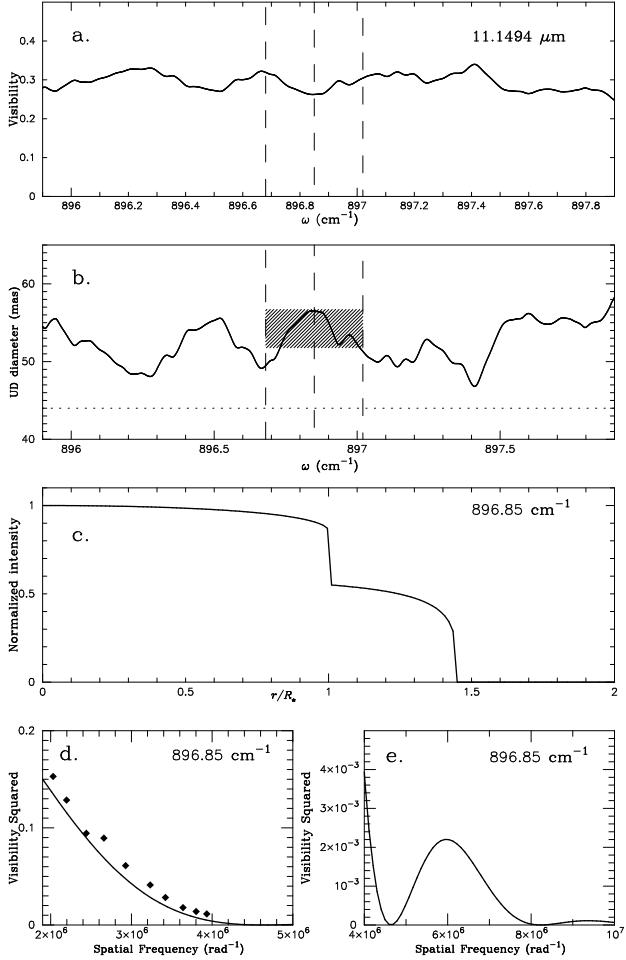}}}
\end{center}
\caption{{\bf a:} Visibility calculated from the best-fit model for 
\mbox{$\alpha$~Ori}\ in the region around 11.1494~\mbox{$\mu$m}.  
The visibility shown here is calculated for a projected baseline 
length of 30~m.  
{\bf b:} Uniform disk diameter calculated from the spectrally 
convolved visibility shown in {\bf a} for a projected baseline 
length of 30~m. 
In {\bf a} and {\bf b}, two bandpasses used by WHT03a are marked 
with the dashed lines.  
The bandpasses used for the other observations of 
\mbox{$\alpha$~Ori}\ lie between those shown (see WHT03a).  
The range of the diameters measured in these bandpasses is shown 
as the hatched region.  
The dotted line represents the photospheric angular diameter 
adopted in the calculation. 
A flux contribution of 44\% from the dust shell is included  
in the calculations of the model visibility, but not in 
the computation of the uniform disk diameter.  
Both plots ({\bf a} and {\bf b}) are redshifted by 0.117~\mbox{cm$^{-1}$}\ 
with respect to the rest 
wavenumber, which corresponds to a radial velocity of 39~\mbox{km s$^{-1}$}, 
to match the observed spectra shown in Fig.~\ref{alfOriSp11mu}. 
{\bf c:} The normalized intensity profile at 896.85~\mbox{cm$^{-1}$}, 
which is spectrally convolved with the same top-hat function as used 
in the calculation of the spectrally convolved visibility shown in {\bf a}.  
{\bf d:} The spectrally convolved visibility squared at 
896.85~\mbox{cm$^{-1}$}\ 
(solid line), plotted together with the visibility points observed 
by WHT03a (filled diamonds).  
{\bf e:} The same spectrally convolved visibility squared as in {\bf d}, 
but at higher spatial frequencies. 
}
\label{alfOriVis11mu}
\end{figure}

The comparison between the synthetic spectra and those observed 
in the 11~\mbox{$\mu$m}, 12~\mbox{$\mu$m}, and 6~\mbox{$\mu$m}\ regions 
demonstrates that the warm water vapor envelope extending to 
$\sim 1.45$~\mbox{$R_{\star}$}\ with a temperature of 2050~K and 
an \mbox{H$_2$O}\ column 
density of $2 \times 10^{20}$~\mbox{cm$^{-2}$}\ can reproduce the observed 
spectra.  
The calculation of the synthetic spectra at 11~\mbox{$\mu$m}\ shows 
that the presence of the warm water vapor envelope can yield a 
featureless, continuum-like spectrum in the bandpasses used 
in the ISI observations by WHT03a.   
If the featureless spectra observed at 11~\mbox{$\mu$m}\ are a result 
of the filling-in effect due to the emission from the outer part of 
the warm water vapor envelope, the angular diameter in this
wavelength region can appear larger than the photospheric
diameter.  This possibility can be examined by computing the 
visibility from the model intensity profile at each wavelength.  

We calculate the intensity profile at a wavenumber interval of 
0.001~\mbox{cm$^{-1}$}, and then from this monochromatic intensity profile, 
the monochromatic visibility is obtained using the Hankel 
transform.  The monochromatic visibility is 
convolved with an appropriate response function which represents 
the spectral resolution of the ISI observations by WHT03a.  
The spectrally convolved visibility is calculated as follows:
\begin{equation}
 V_{\rm conv}(f, \omega) = \int V(f, \omega^{\prime}) \, 
 S(\omega^{\prime} - \omega)\, d\omega^{\prime},
\end{equation}
where $V(f, \omega)$ is the monochromatic visibility at 
the spatial frequency $f$ and 
the wavenumber $\omega$, $V_{\rm conv}(f, \omega)$ is the spectrally 
convolved visibility, and $S(\omega)$ is the spectral response 
function of the bandpass.  In the calculation discussed 
here, $S(\omega)$ is assumed to be a top-hat function with a width of 
0.17~\mbox{cm$^{-1}$}, which is the bandwidth used for the ISI observations. 
We assume that the $K$-band angular diameter represents 
the photospheric diameter.  
Dyck et al. (\cite{dyck92}) and Perrin et al. (\cite{perrin04}) 
derived the $K$-band uniform disk diameter of 43 -- 44~mas for 
\mbox{$\alpha$~Ori}, and we adopt 44~mas as the photospheric diameter.  
Once the visibility function resulting from the 
stellar disk and the warm \mbox{H$_2$O}\ envelope 
(not including the dust shell) 
is calculated, the uniform disk diameter at each wavenumber can be 
derived for a given baseline length.  We compute the uniform disk 
diameter for a baseline length of 30~m, which is the mean of the 
baseline lengths used in the ISI observations by WHT03a.  

Figures~\ref{alfOriVis11mu}a and \ref{alfOriVis11mu}b 
show the calculated visibility and 
the uniform disk diameter of the best-fit model for \mbox{$\alpha$~Ori}\ 
in the region around 11.1494~\mbox{$\mu$m}.  
As mentioned in Sect.~\ref{sect_intro}, the 
presence of the extended dust shell lowers the visibility by an 
amount equal to the fraction of the flux contribution of the 
dust shell.  Therefore, the visibility resulting from the stellar disk 
and the warm \mbox{H$_2$O}\ envelope is lowered by a factor of 0.56, which 
accounts for the flux contribution of the stellar disk in this 
wavelength region.  Note, however, that the uniform disk 
diameter is computed from the visibility excluding the dust shell, 
because the effect of the presence of the dust shell is already 
taken into account in the determination of the uniform disk diameters 
by W00 and WHT03a.  Therefore, the uniform disk 
diameter shown in Fig.~\ref{alfOriVis11mu} can readily be compared 
with those observationally derived by W00 and WHT03a.  
As Fig.~\ref{alfOriVis11mu}b illustrates, the uniform disk 
diameter in this spectral region is larger than the photospheric 
diameter of 44~mas (dotted line in the figure), and 
the predicted diameter in the ISI bandpasses between 896.7 and
897.0~\mbox{cm$^{-1}$}\ is in agreement with the result obtained by WHT03a.  

Figure~\ref{alfOriVis11mu}c shows the intensity profile at the center 
of the range of the bandpasses used by WHT03a, 
while Fig.~\ref{alfOriVis11mu}d shows 
the corresponding visibility squared as a function of spatial
frequency, together with the observed values presented in Fig.~1 of WHT03a.  
Fig.~\ref{alfOriVis11mu}d demonstrates that the predicted visibility 
squared is somewhat lower that those observed.  The observed visibility 
squared can be well fitted with a uniform disk diameter of 52.66~mas 
(WHT03a), 
while the predicted uniform disk diameter is 56.6~mas.  However, 
Weiner et al. (\cite{weiner03b}) show that the uniform disk diameter 
measured for \mbox{$\alpha$~Ori}\ fluctuates with an amplitude of 
$\pm \sim 1.5$~mas, 
and given this temporal variation of the angular diameter of 
\mbox{$\alpha$~Ori}\ 
on the one hand and the simplicity of our model on the other hand, 
the agreement between the observed and predicted visibility squared 
can be regarded as fair.  
Figure~\ref{alfOriVis11mu}e shows the same predicted visibility squared 
at higher spatial frequencies, which correspond to a projected baseline 
as long as $\sim 100$~m.  Observations with such long baselines may be 
realized by ISI in the near future, and the visibility shape at such 
high spatial frequencies will be useful for further examining the 
model for the warm water vapor envelope.  

\begin{figure}
\begin{center}
\resizebox{8.5cm}{!}{\rotatebox{0}{\includegraphics{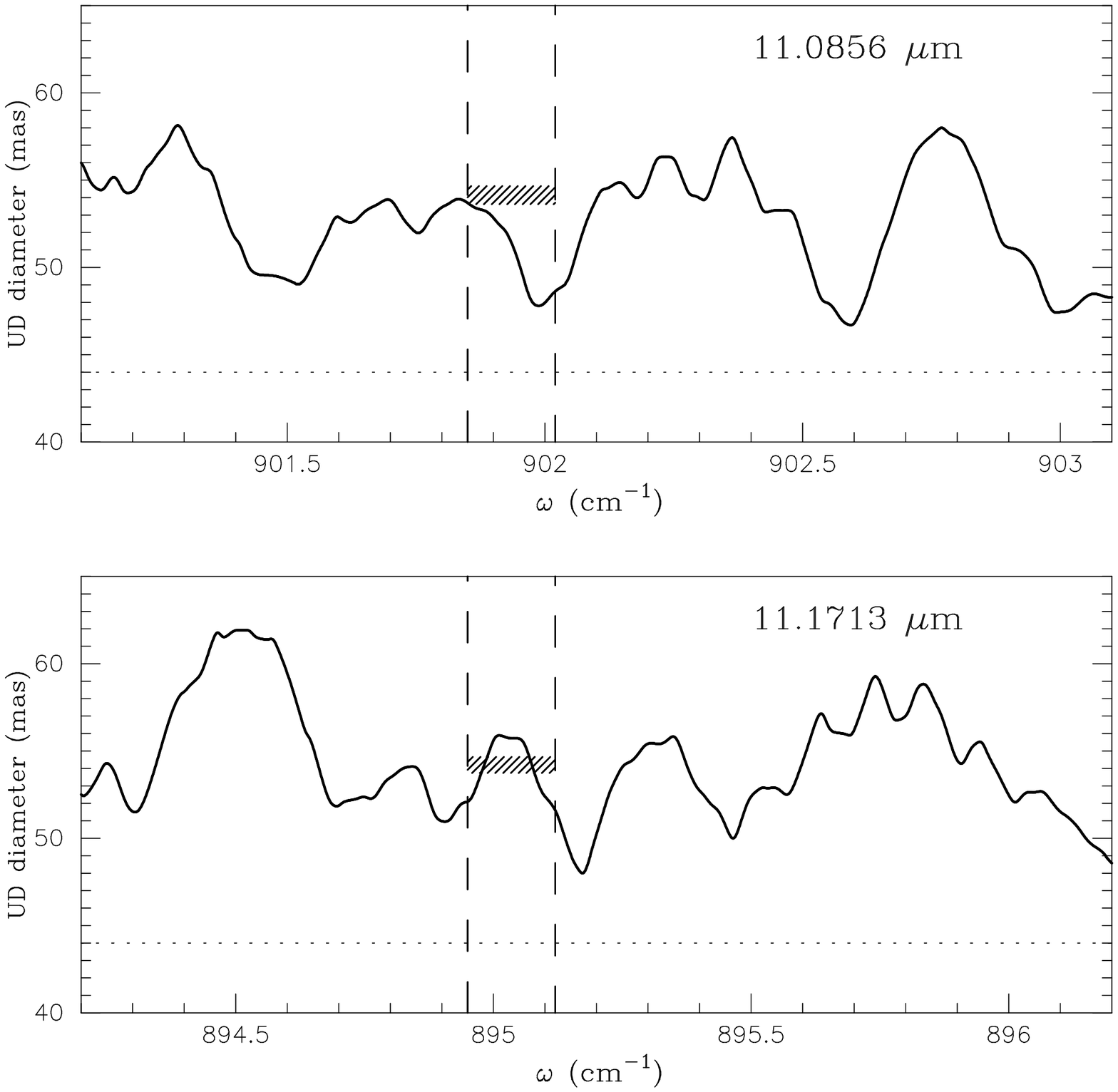}}}
\end{center}
\caption{Uniform disk diameter around 11.0856~\mbox{$\mu$m}\ (upper panel) 
and 11.1713~\mbox{$\mu$m}\ (lower panel) 
calculated from the best-fit model for \mbox{$\alpha$~Ori}.  
Both plots are redshifted by 0.117~\mbox{cm$^{-1}$}\ in wavenumber, 
which corresponds to a radial velocity of 39~\mbox{km s$^{-1}$}, 
to match the observed spectra shown in Fig.~\ref{alfOriSp11mu}. 
The bandpasses used by WHT03a are marked with the dashed lines, 
and the ranges of the angular diameters measured by WHT03a are represented 
as the hatched regions.  
The dotted lines represent the photospheric angular diameter 
adopted in the calculation. 
}
\label{alfOriUD11mu}
\end{figure}

We perform the same calculation for the other two spectral 
regions observed by WHT03a.  Figure~\ref{alfOriUD11mu} shows 
the predicted uniform disk diameter in the regions around 
11.0856~\mbox{$\mu$m}\ and 11.1713~\mbox{$\mu$m}.  
In the 11.0856~\mbox{$\mu$m}\ region, 
the predicted uniform disk diameter ranges from 48 to 53~mas 
within the bandpass used by WHT03a, which are marked with the 
dashed lines in the figure.  These predicted diameters are 
systematically lower than the observed value of $54.14 \pm 0.52$~mas. 
Given the above mentioned temporal variation of the angular diameter 
of \mbox{$\alpha$~Ori}, however, this slight discrepancy is not regarded as 
serious disagreement.  
The predicted diameter in the ISI bandpass in the 11.1713~\mbox{$\mu$m}\ 
region ranges from 51 to 56~mas (lower panel), which is in 
good agreement with the observed value of $54.20 \pm 0.46$~mas. 
Thus, our simple model of the warm 
\mbox{H$_2$O}\ envelope can reasonably explain the observed increase of the 
angular diameter of \mbox{$\alpha$~Ori}\ and the spectra in the 
near-infrared (6 -- 7~\mbox{$\mu$m}) and in the mid-infrared 
(11~\mbox{$\mu$m}\ and 12~\mbox{$\mu$m}). 

\subsection{Uniform disk diameter in the $K$ and \mbox{$L^{\prime}$}\ bands}

While the emission from the warm \mbox{H$_2$O}\ envelope in 
\mbox{$\alpha$~Ori}\ can explain the mid- and near-infrared spectra 
as well as the 
angular diameter measured with the narrow bandpasses at 11~\mbox{$\mu$m}, 
a question may arise: if such a dense \mbox{H$_2$O}\ envelope is 
present, can it affect the angular diameter in the near-infrared?  
T00a demonstrates that the absorption feature at 1.9~\mbox{$\mu$m}\ observed 
in \mbox{$\alpha$~Ori}\ can be explained by the warm \mbox{H$_2$O}\ 
envelope, confirming 
that the presence of the warm molecular envelope can be detected 
by detailed stellar spectroscopy in the near-infrared.  
If the angular diameter measured in the near-infrared were also 
significantly affected by the warm molecular envelope, 
our assumption that the $K$-band angular 
diameter of \mbox{$\alpha$~Ori}\ represents the photospheric diameter could 
not be justified, and hence it might not be concluded that the 
warm \mbox{H$_2$O}\ envelope is responsible for the increase of the apparent 
diameter from the near-infrared to the mid-infrared.  
Moreover, the \mbox{$L^{\prime}$}-band angular diameter of 
\mbox{$\alpha$~Ori}\ 
measured by Chagnon et al. (\cite{chagnon02}) is 42 -- 43~mas, 
which is very close to the $K$-band diameter.  
This suggests that the effect of the warm molecular envelope 
on the \mbox{$L^{\prime}$}-band diameter should not be prominent.  

We examine the effect of the warm \mbox{H$_2$O}\ envelope on the 
angular diameter measured in the near-infrared by performing the same 
calculations as in the previous subsection, 
for the $K$ and \mbox{$L^{\prime}$}\ bands.  
Only \mbox{H$_2$O}\ lines were included in the calculation, and we approximate 
the $K$- and \mbox{$L^{\prime}$}-band filter response functions with top-hat 
functions centered at 2.15~\mbox{$\mu$m}\ with 
$\Delta \lambda$ = 0.5~\mbox{$\mu$m}\ 
and at 3.8~\mbox{$\mu$m}\ with $\Delta \lambda$ = 0.6~\mbox{$\mu$m}, 
respectively.  
The uniform disk diameters predicted in the $K$ and \mbox{$L^{\prime}$}\ 
bands have turned out to be 44.5~mas and 45.6~mas, respectively, and
these values are very close to the 44~mas which we adopted as the 
photospheric angular diameter.  The predicted angular 
diameter in the \mbox{$L^{\prime}$}\ band is also in agreement with the values 
derived by Chagnon et al. (\cite{chagnon02}).  

The optical depth due to the \mbox{H$_2$O}\ lines in the $K$ and 
\mbox{$L^{\prime}$}\ bands 
ranges from $\sim 0.01$ to $\sim 1$ for the gas temperature and 
the \mbox{H$_2$O}\ column density of the best-fit model 
(\mbox{$T_{\rm mol}$}\ = 2050~K and 
\mbox{$N_{\rm col}$}\ = $2 \times 10^{20}$~\mbox{cm$^{-2}$}), 
while Fig.~\ref{alfOriIntensSp11mu} shows that the optical depth in 
the 11.1494 region mostly ranges from $\sim 0.01$ to $\sim 1$.  
It means that the \mbox{H$_2$O}\ envelope is not yet 
totally optically thick, and emission from the star can also be observed.  
In such a case, the effect of emission from the warm \mbox{H$_2$O}\ 
envelope on the angular diameter is not solely governed by 
emission from the envelope, but by the intensity ratio 
between the star and the envelope.  With the emission from the star 
and the warm \mbox{H$_2$O}\ envelope represented with blackbodies of 
\mbox{$T_{\rm eff}$}\ = 3600~K and \mbox{$T_{\rm mol}$}\ = 2050~K, 
respectively, the intensity ratio is 
expressed as 
$B_{\lambda}(\mbox{$T_{\rm mol}$})$/$B_{\lambda}(\mbox{$T_{\rm
eff}$})$. 
This intensity 
ratio is small in the near-infrared, but increases toward 
longer wavelengths.  Therefore, the effect of the warm \mbox{H$_2$O}\ envelope 
on the angular size is much less prominent in the $K$ and 
\mbox{$L^{\prime}$}\ bands 
than in the 11~\mbox{$\mu$m}\ region.  
In the $K$-band spectrum of \mbox{$\alpha$~Ori}, the absorption due to 
CO and CN 
originating in the photosphere is present.  However, the effects of 
these molecular absorption features on the $K$-band angular diameter 
are expected to be minor, 
because the geometrical thickness of the photosphere of \mbox{$\alpha$~Ori}\ 
predicted by the classical hydrostatic model is only 9\% 
of the stellar continuum radius (see T00a), and is much 
smaller than the radius of the warm \mbox{H$_2$O}\ envelope of 
$\sim 1.45$~\mbox{$R_{\star}$}.

\section{Comparison with the observed spectra and angular diameters:
\mbox{\boldmath $\alpha$}~Her}

\begin{figure}
\begin{center}
\resizebox{8.5cm}{!}{\rotatebox{-90}{\includegraphics{0668fig8.ps}}}
\end{center}
\caption{Spectra in the region around 11.1494~\mbox{$\mu$m}.  
The dots represent the observed spectrum of \mbox{$\alpha$~Her}, 
while the solid line represents the synthetic spectrum from 
the best-fit model.  The synthetic spectrum is convolved with 
a Gaussian with a FWHM of 0.037~\mbox{cm$^{-1}$}, and blueshifted 
by 0.084~\mbox{cm$^{-1}$}\ in wavenumber, 
which corresponds to a radial velocity of $-28$~\mbox{km s$^{-1}$}\ 
(see WHT03a), 
to match the observation. 
The positions of the \mbox{H$_2$O}\ lines whose intensity at 2000~K is 
stronger 
than $3 \times 10^{-23}$~cm molecule$^{-1}$ are marked with ticks.  
These line positions are taken from the HITEMP database, and 
are also blueshifted by 28~\mbox{km s$^{-1}$}\ with respect to the 
rest wavenumber. 
}
\label{alfHerSp11mu}
\end{figure}

\begin{figure}
\begin{center}
\resizebox{8.5cm}{!}{\rotatebox{-90}{\includegraphics{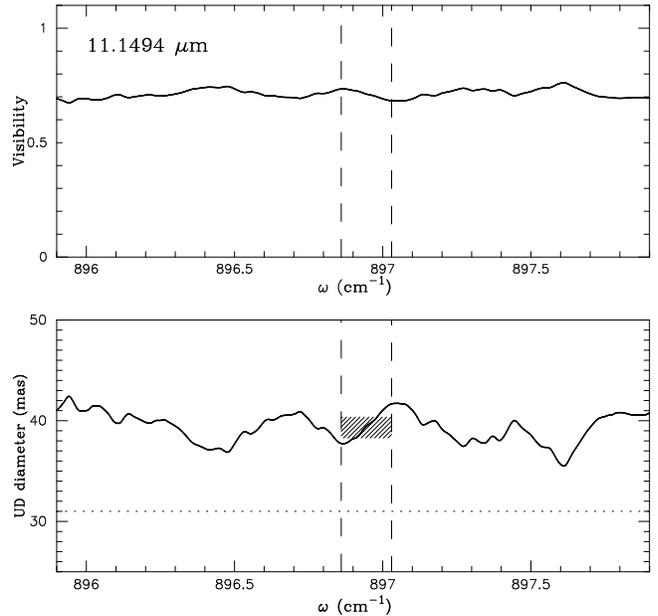}}}
\end{center}
\caption{
Visibility and uniform disk diameter 
calculated from the best-fit model for \mbox{$\alpha$~Her}\ 
in the region around 11.1494~\mbox{$\mu$m}.  
The visibility and the uniform disk diameter shown are calculated 
for a projected baseline length of 30~m.  
No flux contribution of the dust shell is assumed  
in the calculation of the model visibility and uniform disk diameter.  
The dotted line represents the adopted photospheric angular diameter, 
while the range of the angular diameter measured by WHT03a 
is shown as the hatched region. 
Both plots are blueshifted by 0.084~\mbox{cm$^{-1}$}\ with respect to the 
rest wavenumber to match the observation. 
}
\label{alfHerVis11mu}
\end{figure}

For \mbox{$\alpha$~Her}, we compare the 
11~\mbox{$\mu$m}\ spectrum presented in WHT03a 
and the angular diameter measured in this spectral range. 
With an effective temperature of 3200~K adopted for \mbox{$\alpha$~Her}, 
we search for the combination of the 
input parameters which can best reproduce the spectra and 
the angular diameter observed in the 11~\mbox{$\mu$m}\ region.  
The parameters of the best-fit model are found to be 
\mbox{$T_{\rm mol}$}\ = 2000~K, \mbox{$R_{\rm mol}$}\ = 
1.4~\mbox{$R_{\star}$}, and 
$N_{\rm col} = 7 \times 10^{20}$~\mbox{cm$^{-2}$}.  
We estimate that the uncertainties of 
the gas temperature, the radius, and the column density of \mbox{H$_2$O}\ 
are $\pm 100$~K, $\pm 0.1$~\mbox{$R_{\star}$}, and a factor of $\sim 2$, 
respectively.  

Figure~\ref{alfHerSp11mu} shows a comparison between the observed 
spectrum (dots) and the calculated one from the best-fit model 
(solid line).  As in the case of 
\mbox{$\alpha$~Ori}, the synthetic spectrum is convolved with a Gaussian 
which represents the effects of the instrument as well as of the 
macroturbulent velocity in the atmosphere of \mbox{$\alpha$~Her}.  
Tsuji (\cite{tsuji88}) analyzed the high-resolution spectra of 
the CO first overtone bands of M giants with the use of 
photospheric models, and detected an absorption excess in the 
low excitation CO lines, which he concluded originates in 
the warm molecular envelope.  
Based on the analysis of this absorption excess, 
Tsuji (\cite{tsuji88}) found that the turbulent velocity 
in the warm molecular envelope of the M giants in his sample, 
including \mbox{$\alpha$~Her}, 
can be as large as 10~\mbox{km s$^{-1}$}.  Such a large turbulent velocity is 
in fact observed in the 12~\mbox{$\mu$m}\ spectrum of 
\mbox{$\alpha$~Ori}, as mentioned in 
Sect.~\ref{sect_alfOri11mu}.  In the present work, we tentatively 
assume the same macroturbulent velocity in \mbox{$\alpha$~Her}\ 
as in \mbox{$\alpha$~Ori}, 
that is, 12~\mbox{km s$^{-1}$}.  
Since \mbox{$\alpha$~Her}\ 
shows no dust emission feature in the $N$ band (Monnier et al. 
\cite{monnier98}), no dilution effect due to dust emission is 
included in the calculation for \mbox{$\alpha$~Her}.  
A glance of Fig.~\ref{alfHerSp11mu} reveals that 
the model can reproduce the observed continuum-like
spectrum nearly free from salient features and, to some extent, 
the depths of the fine spectral features. 

We calculate the visibility and the uniform disk diameter in 
this spectral region.  In the 
calculation of the visibility, we assume that the photospheric 
angular diameter can be represented by the $K$-band angular diameter.  
Benson et al. (\cite{benson91}) and Perrin et al. (\cite{perrin04}) 
derived  $K$-band angular diameters of 31 -- 32~mas for \mbox{$\alpha$~Her}, 
and we adopt an angular diameter of 31~mas as the 
photospheric angular diameter in our calculation.  
The uniform disk diameter is derived 
from the calculated visibility for a baseline length of 30~m.  
Figure~\ref{alfHerVis11mu} shows that the uniform disk diameter is 
much larger than the photospheric diameter of 31~mas in the spectral 
region at issue.  In the bandpass 
used in the diameter measurements by WHT03a, which are marked with the 
dashed lines in the figure, the angular diameter ranges from 38 to 42~mas, 
which is in agreement with the $39.32 \pm 1.04$~mas measured by WHT03a.  

We also examine the effect of the warm \mbox{H$_2$O}\ envelope on the 
diameter in the $K$ and \mbox{$L^{\prime}$}\ bands.  
The uniform disk diameters 
in the $K$ and \mbox{$L^{\prime}$}\ bands predicted from the 
best-fit model for 
\mbox{$\alpha$~Her}\ are 33.3~mas and 33.9~mas, respectively, and 
the predicted \mbox{$L^{\prime}$}-band diameter is in agreement with the 
observed diameter of $31.04 \pm 0.26$~mas derived by 
Chagnon et al. (\cite{chagnon02}).  
This confirms that the 
$K$-band and \mbox{$L^{\prime}$}-band diameters are not significantly affected 
by the presence of the warm \mbox{H$_2$O}\ envelope, and that the use of 
the $K$-band diameter as the photospheric diameter is reasonable. 
It should also be noted that the parameters of the best-fit model 
for \mbox{$\alpha$~Her}\ 
are in rough agreement with those derived by Tsuji (\cite{tsuji88}) based 
on the analysis of the absorption excess in the low excitation lines 
of the CO first overtone bands.

\section{Discussion}

We have shown that the increase of the angular diameters 
from the near-infrared to 11~\mbox{$\mu$m}\ observed in the supergiants 
\mbox{$\alpha$~Ori}\ and \mbox{$\alpha$~Her}\ can be explained by 
the presence of a warm \mbox{H$_2$O}\ 
envelope, and that dense \mbox{H$_2$O}\ gas with a temperature of 
$\sim 2000$~K 
extends to 1.4 -- 1.5~\mbox{$R_{\star}$}.  These results are in good agreement 
with those recently obtained by Perrin et al. (\cite{perrin04}), 
who could reproduce the angular diameter of \mbox{$\alpha$~Ori}\ 
measured in the 
$K$ and \mbox{$L^{\prime}$}\ bands as well as that at 11.15~\mbox{$\mu$m}, 
using a spherical gaseous envelope model without a line-by-line 
calculation of molecular opacities.  They derived the 
temperature and the radius of such a gaseous envelope of 
\mbox{$\alpha$~Ori}\ to 
be 2055~K and 1.33~\mbox{$R_{\star}$}, respectively, with optical depths of 
0.06, 0.026, and 2.33 in the $K$ band, \mbox{$L^{\prime}$}\ band, 
and at 11.15~\mbox{$\mu$m}, 
respectively.  Since our model predicts the optical depth as a function 
of wavenumber, we average the optical depth predicted from our 
best-fit model for \mbox{$\alpha$~Ori}\ in the $K$ and 
\mbox{$L^{\prime}$}\ bands.  
The averaged optical depths of the warm water vapor envelope predicted 
from our model 
are 0.05 and 0.026 for the $K$ and \mbox{$L^{\prime}$}\ bands, respectively, 
which is in good agreement with the above values derived by 
Perrin et al. (\cite{perrin04}).  For the optical depth in the
11.15~\mbox{$\mu$m}\ region, Fig.~\ref{alfOriIntensSp11mu} shows that 
the optical depth predicted from our best-fit model for \mbox{$\alpha$~Ori}\ 
ranges from $\sim 0.01$ up to 10, which is also in rough agreement 
with the value derived by Perrin et al. (\cite{perrin04}).  

From UV observations, M supergiants are known to have chromospheres 
as hot as $T_e \sim 8000$~K.  
Gilliland \& Duprees (\cite{gilliland96}) obtained the first image 
of \mbox{$\alpha$~Ori}\ at 2550~\AA\ with the Hubble Space Telescope, and 
found that the chromospheric extension of \mbox{$\alpha$~Ori}\ is 
about 3 times 
as large as the size measured in the $K$ band.  However, the VLA 
observations of \mbox{$\alpha$~Ori}\ by Lim et al. (\cite{lim98}) revealed 
the presence of 
cooler gas with temperatures of 1300 -- 3400~K at 2 -- 7~\mbox{$R_{\star}$}. 
Lim et al. (\cite{lim98}) conclude that the hot chromosphere and 
the cool gas coexist, but with the latter being the more dominant component.  
Harper et al. (\cite{harper01}) constructed a semiempirical model to 
explain these VLA observations, and their one-dimensional model has 
a temperature distribution which first decreases outward from the 
photosphere, and 
rises to the maximum of $\sim 3800$~K at $\sim 1.45$~\mbox{$R_{\star}$}, 
and then decreases again.  
While the temperature and the radius of the warm \mbox{H$_2$O}\ envelope 
we derived for \mbox{$\alpha$~Ori}\ ($\sim 2050$~K at 
$\sim 1.45$~\mbox{$R_{\star}$}) 
seem to be in conflict with this semiempirical model, 
it should be noted that Harper et al. (\cite{harper01}) 
suggest that their one-dimensional 
model might represent the average of an inhomogeneous 
structure, where the cool gas ($\sim 2000$~K) and the hot plasma 
($\sim 8000$ -- 10\,000~K) coexist at $\sim 1.45$~\mbox{$R_{\star}$}. 
Then it is implied that the warm \mbox{H$_2$O}\ envelope which we modeled 
in the present work may be part of the cool 
component detected by the VLA observations and modeled by 
Harper et al. (\cite{harper01}). 

The increase of the angular diameter is detected not only in 
supergiants but also in Miras and non-Mira M giants.  
Weiner et al. (\cite{weiner03b}) as well as W00 and WHT03a 
observed the Mira variables \object{$o$~Cet}, \object{R~Leo}, 
and \object{$\chi$~Cyg} with 
the same observational technique as applied to \mbox{$\alpha$~Ori}\ 
and \mbox{$\alpha$~Her}, and 
found that the angular diameters of these Mira variables are roughly 
twice as large as those measured in the $K$ band. 
As in the cases of \mbox{$\alpha$~Ori}\ and \mbox{$\alpha$~Her}, 
they used bandpasses which 
appear not to be contaminated 
by \mbox{H$_2$O}\ or other spectral lines.  However, as we have shown above, 
the interpretation of the mid-infrared spectra of late-type 
(super)giants is complicated by flux contribution from the extended 
outer atmosphere.  
The increase of diameter is detected not 
only in the mid-infrared, but also in the \mbox{$L^{\prime}$}\ band.  
Menneson et al. (\cite{mennesson02}) show that the \mbox{$L^{\prime}$}-band 
angular diameters of Mira variables as well as semiregular M giants 
are by a factor of 1.2 -- 2.0 larger than the \mbox{$K^{\prime}$}-band 
diameters.  
The increase of the angular diameters from the \mbox{$K^{\prime}$}-band to 
the \mbox{$L^{\prime}$}-band and the 11~\mbox{$\mu$m}\ region observed 
in the Mira variables and semiregular M giants 
may also be explained by the warm molecular envelope, whose presence 
in these classes of objects is detected by the analyses of 
infrared molecular spectra (e.g., Tsuji et al. \cite{tsuji97}, 
Yamamura et al.~\cite{yamamura99}, Cami et al. \cite{cami00}, 
Matsuura et al.~\cite{matsuura02}). 
This possibility will be further studied in a forthcoming paper 
(Ohnaka \cite{ohnaka04}). 

Although our ad hoc model for the warm molecular envelope can 
reproduce the observed spectra as well as the angular diameter 
increase from the near-infrared to the mid-infrared, 
it is not a unique solution.  
The hypothesis of the warm molecular envelope should be further 
examined by comparing with spectroscopic and interferometric 
data in other wavelength regions.  For example, interferometric 
observations in many more bandpasses in the 11~\mbox{$\mu$m}\ region 
would provide a more complete picture of the wavelength dependence 
of the angular diameter, which would be a further constraint for 
modeling the warm molecular envelope.  Interferometry with even higher 
spectral resolution would also be very useful 
for testing the hypothesis proposed in the present work. 
Even if the basic picture 
of the warm water vapor envelope is correct, our simple model is not 
sufficient to understand detailed physical properties of the 
warm molecular envelope such as temperature and density 
distributions, and therefore, the physical parameters we 
derived above should be regarded as representative values 
of the real molecule forming region.  

Furthermore, the physical 
mechanism responsible for the formation of the warm molecular 
envelope remains to be answered.  
In the coolest and the most luminous objects such as Mira variables, 
the levitation of the atmosphere due to stellar pulsation may 
lead to a density enhancement in the outer atmosphere, and hence 
creating an environment favorable for molecules to form, 
as Helling \& Winters (\cite{helling01}) discuss for carbon-rich 
objects.  However, it is not yet clear whether such a mechanism 
can operate in M (super)giants with higher effective temperatures and 
much smaller variability amplitudes.  
For example, Tsuji et al. (\cite{tsuji97}), 
Matsuura et al. (\cite{matsuura99}), and Tsuji (\cite{tsuji01}) 
detected water vapor 
in the near-infrared spectra of early M giants and a late K giant, 
whose photospheres had been deemed to be too hot for water vapor 
to form.  
Ryde et al. (\cite{ryde02}) detected \mbox{H$_2$O}\ pure-rotation lines at 
11 -- 12~\mbox{$\mu$m}\ even in the K1.5III giant \object{$\alpha$~Boo}.  
They argue, however, that these \mbox{H$_2$O}\ lines do not originate 
in the warm 
molecular envelope, but in the outer layers of the photosphere where 
temperature may deviate from that predicted by classical, hydrostatic 
photospheric models, although the mechanism responsible for such 
a deviation remains ambiguous.  
In any case, the physical properties of the 
region between the upper photosphere and the expanding, cold 
circumstellar envelope have been increasingly probed with various 
observational techniques.  The understanding of physical 
processes in operation should also be pursued from theoretical 
point of view to explain the rather common occurrence of 
the warm molecular envelope, or exactly speaking, the component 
which cannot be explained in the framework of classical, hydrostatic 
photospheric models, in a wide range of late-type stars from 
Mira variables to M (super)giants and K giants.

\section{Concluding remarks}

Our simple model of the warm \mbox{H$_2$O}\ envelope can simultaneously 
reproduce the spectra and the angular diameters of \mbox{$\alpha$~Ori}\ 
and \mbox{$\alpha$~Her}\ 
obtained at 11~\mbox{$\mu$m}.  For \mbox{$\alpha$~Ori}, we have also 
shown that the spectra obtained at 12~\mbox{$\mu$m}\ as well as 
at 6 -- 7~\mbox{$\mu$m}\ can be fairly reproduced by this model. 
The continuum-like spectra of \mbox{$\alpha$~Ori}\ and \mbox{$\alpha$~Her}\ 
observed in the 11~\mbox{$\mu$m}\ region 
can be interpreted as a result of the filling-in due to emission 
from the outer part of the warm \mbox{H$_2$O}\ envelope.  
Although the observed featureless 
11~\mbox{$\mu$m}\ spectra do not show any hint of the presence of the warm 
\mbox{H$_2$O}\ envelope, it manifests itself as an increase of the 
angular diameter.  
Our model can reproduce the increase of the 
angular diameters observed for \mbox{$\alpha$~Ori}\ and 
\mbox{$\alpha$~Her}\ from the $K$ band 
to the 11~\mbox{$\mu$m}\ region.  
For \mbox{$\alpha$~Ori}, the gas temperature, 
the \mbox{H$_2$O}\ column density, 
and the radius of this \mbox{H$_2$O}\ envelope 
were derived to be 2050~K, $2 \times 10^{20}$~\mbox{cm$^{-2}$}, 
and 1.45~\mbox{$R_{\star}$}, 
respectively.   For \mbox{$\alpha$~Her}, we derived 
a water vapor gas temperature of 2000~K, a column density of 
$7 \, \times \, 10^{20}$~\mbox{cm$^{-2}$}, and a radius of
1.4~\mbox{$R_{\star}$}. 

\begin{acknowledgement}
The author would like to thank Prof.~T.~Tsuji and Dr.~T.~Driebe 
as well as the anonymous referee for valuable comments. 
\end{acknowledgement}

\end{document}